\documentclass[preprint,preprintnumbers,showpacs,prl]{revtex4}
\usepackage[english]{babel}
\usepackage{amsmath}

\usepackage{latexsym}

\usepackage{epsfig}

\newcommand{\be}{\begin{equation}}
\newcommand{\ee}{\end{equation}}
\newcommand{\bea}{\begin{eqnarray}}
\newcommand{\eea}{\end{eqnarray}}
\newcommand{\cn}{\mbox{\rm cn}\,}
\newcommand{\sn}{\mbox{\rm sn}\,}
\newcommand{\dn}{\mbox{\rm dn}\,}
\newcommand{\prt}{\partial}
\newcommand{\rgl}{\rangle}
\newcommand{\lgl}{\langle}

\begin{document}

\title{Sensitivity of ray paths to initial conditions}
\author{A. Iomin$^{a}$ and G.M. Zaslavsky$^{b,c}$ }
\affiliation{$^a$ Department of Physics, Technion, Haifa, 32000, Israel.  \\
$^b$ Courant Institute of Mathematical Sciences, \\
New York University, 251 Mercer Str., New York, NY 10012 \\
$^c$  Department of Physics, New York University, \\
2-4 Washington Place, New York, NY 10003. }

\begin{abstract}
Using a parabolic equation, we consider ray propagation in a wave\-guide
with
the sound speed profile that corresponds to the dynamics of a nonlinear
oscillator. An analytical consideration of the dependence of the
travel  time on the initial conditions is presented.
Using an exactly solvable model and the path integral representation
of the travel time, we explain the step-like behavior of the travel time
\(T\)
as a function of the starting momentum \(p_0\) (related to the starting
ray
grazing angle \(\chi_0\) by \(p_0=\tan\chi_0\)).
A periodic perturbation of the waveguide along the range leads to  wave
and ray chaos. We explain an inhomogeneity of distribution of the chaotic
ray travel times, which has obvious maxima.
These maxima lead to the clustering of rays and each maximum relates to a
ray identifier, {\em i.e.} to the number of ray semi--cycles along the ray
path.


\end{abstract}

\maketitle

\section{Introduction}

The ray tracing has demonstrated that early parts of the time-fronts, {\em
i.e.} ray arrivals in time-depth plane,  exhibit surprisingly regular
structures simultaneously with sensitivity to
the media inhomogeneities that give rise to ray chaos
\cite{SFW97,BV98,group2,AET1,AET2}. Even at very long ranges, up to a few
thousand km, steep chaotic rays can form remarkably
stable segments of the time-front, which are very close to their
prototypes in the unperturbed waveguide. It turns out that each stable
segment in the perturbed waveguide, like its counterpart in the unperturbed
one, is formed by rays with equal identifiers, {\em i.e.} the number of ray
semi--cycles.

This phenomenon can be observed in the travel times of the so-called
eigenrays,
i.e. rays passing through a fixed observation point. In Ref. \cite{TT96}
(see also Refs. \cite{PGJ91,ST96}) it has been demonstrated that the
travel times
of chaotic eigenrays usually come in clusters with small time spreads
centered at arrivals of unperturbed eigenrays. Although the rays
that form a cluster have the same identifier, i.e. the same topology,
this does
not imply that their trajectories follow close paths. On the contrary,
chaotic eigenrays contributing to the given cluster may significantly deviate
from each other and from the unperturbed eigenray with the same identifier
\cite{group2}.
These results suggest that while the travel time of chaotic rays is a random
function of starting parameters, it is much more predictable as a function
of its identifier and the trajectory endpoints, and it also
relates to the dependence of the travel time \(T\) on the starting
momentum \(p_0\) \cite{SVZ02}. Interesting, even puzzling features, when
observed numerically, reveal a step-like behavior of \(T\)
as a function of the initial momentum that  forms the so-called
``shelves'' for ray propagation in a range-independent waveguide.
In the case when rays propagate in a range-dependent waveguide
so-called, ray chaos takes place, and an important characteristic
is a ray travel time distribution, which exhibits fairly inhomogeneous
features with obvious maxima \cite{SVZ02}.

In this paper we study analytically the dependence of the travel time on
the initial conditions in the framework of an exact solvable model.
We consider ray propagation in a waveguide with
the sound speed profile which corresponds to the dynamics of a quartic
oscillator. Therefore, studying this model,
we explain the step-like behavior of the travel time \(T\) as a function
of the starting momentum \(p_0\) (related to the starting ray grazing
angle
\(\chi_0\) by \(p_0=\tan\chi_0\)). For the case when ray chaos takes place
due to a range dependent perturbation, we also explain the inhomogeneity
of the ray travel time distribution which has obvious maxima.
These maxima lead to the clustering of rays, and each maximum can be
related to the corresponding identifier of the cluster of rays.

The paper is organized as follows. In Sec. 2 we give a brief description
of ray theory in the small--angle approximation. Explicit expressions
for the Hamiltonian determining the ray motion and for the ray travel
time are presented. An overview of the numerical results on the ray travel
time, obtained in \cite{SVZ02}, will be presented and explained in
the present paper.
An exact solution for a simplified speed profile
corresponding to the quartic oscillator is considered in Sec. 3.
An exact expression for the corresponding classical action as a function
of the initial conditions is presented. A detailed analytical analysis of
the step-like functional dependence of \(T\) on \(p_0\) is performed as well.
In Sec. 4 the maxima of the distribution function for the ray travel time
are found for the integrable quartic oscillator in the presence of
a perturbation. This analysis corresponds to the semiclassical consideration
of an amplitude of the wave function in the momentum representation.
The conclusion is presented in Sec. 5. Some details of calculations
related to the step-function are presented in Appendices A-C.

\section{Ray travel times}
\subsection{Parabolic equation approximation}

Consider a two-dimensional underwater acoustic waveguide with the sound
speed
\( c \) being a function of depth, \( z, \) and range, \( r \). The sound
wave field \( u \) as a function of \( r, \) \( z \), and time, \( t \),
may be represented as
\begin{equation}
\label{swu}
u(r,z,t)=\int d\omega \, \tilde{u}(r,z,\omega )\, e^{-i\omega t},
\end{equation}
where the Fourier components \( \tilde{u} \) are governed by the Helmholtz
equation (see for example\cite{BL91,JKPS94}):
\begin{equation}
\label{Heq}
\frac{\partial ^{2}\tilde{u}}{\partial r^{2}}+\frac{\partial ^{2}\
\tilde{u}}{\partial z^{2}}+k^{2}n^{2}\tilde{u}=0.
\end{equation}
Here \( k=\omega /c_{0} \) is a wave number, while \( n=c_{0}/c(r,z) \),
is the refractive index and \( c_{0} \) is a reference
sound speed. For the 2D picture, grazing angles are defined as the ratio
between the wave numbers \( k_z \) and \( k_r \):
\( \tan\chi=k_z/k_r \), where \( k=\sqrt{k_z^2+k_r^2} \).
In the small-angle approximation, when sound waves propagate
at small grazing angles with respect to the horizontal, {\em i.e.}
\( k_r\approx k \), the Helmholtz equation may
be approximated by the standard parabolic equation
\cite{BL91,JKPS94,SFW97}.
Present \( \tilde{u} \) as
\begin{equation}
\label{uv}
\tilde{u}(r,z,\omega )=\tilde{v}(r,z,\omega )\, e^{ikr}
\end{equation}
 and substitute this expression into Eq. (\ref{Heq}).
Taking into account that \( \tilde{v} \) is a slowly--varying
function of \(r\) and neglecting
the second derivative of \( \tilde{v} \) with respect to \( r \), we
derive  the parabolic equation
\begin{equation}
\label{Scheq}
2ik\frac{\partial \tilde{v}}{\partial r}+\frac{\partial ^{2}
\tilde{v}}{\partial z^{2}}+k^{2}\left( n^{2}-1\right) \tilde{v}=0.
\end{equation}
This equation coincides formally with the time-dependent Schr\"odinger
equation. In this case the partial derivative with respect to \(z\) is an
analog of the momentum operator, {\em i.e.} \(\hat{p}=-ik^{-1}\prt/\prt z\),
while \( r \) plays the role of time and \( k^{-1} \) associates
with the
Planck constant. In underwater acoustics it is always possible to choose
the reference sound speed \( c_{0} \), such that \( |n-1|\ll 1 \), and
replace   \( 1-n^{2} \) by \( 2(1-n)=2(c(r,z)-c_{0})/c_{0} \).

Since \( r \)
is a time-like variable, the Hamiltonian system formally coincides with
that describing a mechanical particle oscillating in a time-dependent potential
well \(U\)  with the Hamiltonian

\begin{equation}
\label{Ham}
H=\frac{p^{2}}{2}+U(z),
\end{equation}
where
\begin{equation}
\label{U}
U(r,z)=\frac{c(r,z)-c_{0}}{c_{0}}.
\end{equation}
The dimension variable \( p \) is an analog to the mechanical momentum.
It relates to
the ray grazing angle \( \chi  \) by \( p=\tan \chi  \).
The ``potential'' \( U \) in Eq. (\ref{U}) represents
a potential well whose parameters may vary with the range \( r \).

For the point source located at \( r=0 \) and \( z=z_{0} \) we have
\begin{equation}
\label{vtild}
\tilde{v}=\sum _{\nu }A_{\nu }(z,z_{0},r,\omega )\, e^{ikS_{\nu
}(z,z_{0},r)},
\end{equation}
where the sum goes over contributions from all rays connecting the source
and the observation point \( (r,z) \). Such rays are called the
eigenrays\textbf{.}
Here \( S(z,z_{0},r) \) is the eikonal analog to classical action or the
Hamilton principal
function in mechanics -- of the \( \nu  \)-th eigenray. This function is
defined by the integral \cite{LLmech}
\begin{equation}
\label{S}
S=\int \left( pdz-Hdr\right)
\end{equation}
over the ray trajectory from \( (0,z_{0}) \) to \( (r,z) \).

The amplitude \( A(z,z_{0},r) \) is given by \cite{Gutz67}
\begin{equation}
\label{A}
A=C(\omega )\, \sqrt{\left| \frac{\partial ^{2}S}{\partial z\partial
z_{0}}\right| }=C(\omega )\, \sqrt{\frac{1}{\left| \partial z/
\partial p_{0}\right| }},
\end{equation}
where \( C(\omega ) \) is a function determined by the time-dependence of
the radiated signal, and the derivative \( \partial z/\partial p_{0} \) is
taken at the range \( r \).

Substitution of Eqs. (\ref{uv}) and (\ref{vtild}) into Eq. (\ref{swu})
yields
\begin{equation}
\label{vgeom}
u(r,z,t)=\sum _{\nu }\int d\omega \, A_{\nu }(z,z_{0},r,\omega )\, \exp
\left( i\omega \left( \frac{r}{c_{0}}+\frac{1}{c_{0}}S_{\nu }(z,z_{0},r)-t
\right) \right) .
\end{equation}
Each term in this sum represents a sound pulse coming to the observation
point through a particular eigenray. The quantity
\begin{equation}
\label{T}
T=\frac{r}{c_{0}}+\frac{1}{c_{0}}S(z,z_{0},r)
\end{equation}
determines a delay of the pulse relative to an initially radiated signal
and it is called the ray travel time.

\subsection{Numerical results: an overview of \cite{SVZ02}}

Studying the general properties of ray travel times in acoustic
waveguides is equivalent to studying the properties of the principal
Hamiltonian function \( S \) of a mechanical
particle oscillating in a potential well.
Recently the properties of \(S\) have been numerically studied
in \cite{SVZ02}. Hereafter, we refer to this result as SVZ.
The main  numerical results important for the present analysis  are shown
in Figs. 1 and 2, which were taken from SVZ. Both figures present the
travel time dependence on the starting momentum \(p_0\).
Figure 1 demonstrates dependencies of the ray travel time \( T \) on the
starting momentum \( p_{0} \) for two waveguides with different sound speed
profiles, i.e. for two ``potentials'' \( U(z) \):
\begin{equation}
\label{U1}
U_{1}(z)=az^{2}+bz^{4}, ~~~~\mbox{and}~~~~U_{2}(z)=-\gamma z.
\end{equation}
All six curves shown in Fig. 1 present the travel times at a range of 150
km, and each curve corresponds to a particular source depth.
Even though the ``potentials'' \(U_1\) and \(U_2\) are quite different,
both dependencies \( T(p_{0}) \), shown in Figs. 1 and 2, have an
important common feature:
each curve has ``shelves''  where its inclinations with respect to the
horizontal are small. At intervals of starting momenta corresponding to
the ``shelves'', the ray travel time \( T \)
is most stable (least sensitive) with respect to small variations in
\( p_{0} \). The same features occur for the so-called canonical
sound speed profile or the Munk profile \(c(z)=c_M(z)\), widely used in
underwater
acoustics to model wave transmission through a deep ocean
\cite{BL91,FDMWZ79}.
The dependencies \( T(p_{0}) \)  are shown in Fig. 2 which presents the
ray travel times, \( T \), as a function
of  the starting momentum, \( p_{0} \), at range of 2500 km
for a point source located at 1 km depth. A thick solid line graphs the
\( T(p_{0}) \) for the regular ray in the aforementioned range-independent
waveguide. Randomly scattered points correspond to chaotic rays in the
presence of perturbation: \( c(z)=c_M(z)+\delta c(z,r)\). The density
of these points has maxima forming dark horizontal stripes, as is
shown in Fig. 2. It should be pointed out that, while the background
profile \(c(z)\) is realistic, the perturbation \( \delta c(z,r)\) has
been chosen in SVZ to present only a strongly idealized model of
internal-wave-induced sound speed variations \cite{SBT92a,ZA97,V2001}.
Nevertheless, this perturbation causes a chaotic ray behavior
whose properties closely resemble that observed in more realistic
numerical models \cite{SFW97,BV98}.
Ray travel times in the unperturbed waveguide (\( \delta c=0 \))
presented in Fig.2 (thick solid line) have the same
properties typical of range-independent waveguides, namely ``shelves'',
which are similar to those presented in Fig. 1.
Moreover, for the chaotic rays, the stripes of the scattered points are
located at travel times close to that of the ``shelves''
on the unperturbed \( T(p_{0}) \) curves. Note that the unperturbed
``shelves'' may be interpreted as parts of the \( T(p_{0}) \) curve with
the highest density
of points depicting unperturbed arrivals. It has been stated in SVZ that,
under conditions of ray chaos, the positions of maxima of the density of
ray travel times remain relatively stable.
So, these figures express the main puzzling results of the generic
features
of ``shelves'', and our main task in the present work is to explain them.

\section{A quartic oscillator}

As was mentioned above, the general properties of the ray travel time
\(T\)
can be described by the action \(S\) of a mechanical particle in a
potential well. Therefore, the generic features of ``shelves'' for
an unperturbed ray
can be explained in the framework of an analytical description of the
Hamiltonian principal function or the classical action \(S\) (\ref{S}) for
an integrable system with some
potential \(U\) (\ref{U}). Below we consider the oscillating dynamics of a
particle in the potential \(U_1\) of (\ref{U1}).

\subsection{Action}

As the momentum \(p=\tan\chi\) is a dimensionless variable, it is
convenient
to consider the potential in the dimensionless variables as well. Namely
we consider \(\sqrt{2a}z\rightarrow z\) and \(\sqrt{2a}r\rightarrow \omega
r\), while \(b\omega^2/a^2=\lambda\).
Therefore, the dynamical equation for a particle in the potential \(U_1\)
(also called a quartic oscillator) in the new notation is
\begin{equation}\label{Newteq}
\ddot{z}+\omega^2z+\lambda z^3=0.
\end{equation}
Following to SVZ we take a mass \(m=1\). We also use here, formally,
the notation
\(\ddot{z}\equiv d^2z/dr^2\), {\em i.e.} the range \(r\) plays the same
role as a formal time in Hamiltonian dynamics.
This equation can be solved exactly. The solution is chosen in the form
of the Jacobian elliptic function \cite{mizrahi,byrd}
\begin{equation}\label{sol1}
z(r)=Z\cn(\Omega r+\phi_0,\kappa),
\end{equation}
where \(Z\) and \(\phi_0\) are an amplitude and an initial phase of
oscillations correspondingly.
The frequency of the nonlinear oscillations is
\begin{equation}\label{Omega}
\Omega^2=\omega^2+\lambda Z^2
\end{equation}
and the modulus of the elliptic functions is
\begin{equation}\label{kappa}
2\kappa^2=\lambda(Z/\Omega)^2.
\end{equation}
These values are obtained by the straightforward substitution of the
solution
(\ref{sol1}) into (\ref{Newteq}). Following \cite{mizrahi} we take the
modulus \(\kappa\) and the initial phase
\(\phi_0\) to be constants of integration. In this case, the solution
(\ref{sol1}) is
\begin{equation}\label{sol2}
z(r)=Z\cn(\phi,\kappa)=
\left[\frac{2\kappa^2\omega^2}{\lambda(1-2\kappa^2)}\right]^{1/2}
\cn\left[\frac{\omega r}{\sqrt{1-2\kappa^2}}+\phi_0,\kappa\right],
\end{equation}
where \(\kappa\) and \(\phi_0\) are associated with the initial coordinate
\(z_0\) and momentum \(p_0\) as
\begin{equation}\label{rel}
z_0=z(r=0)=Z\cn(\phi_0,\kappa),~~
p_0=\dot{z}(r=0)=-Z\Omega\sn(\phi_0,\kappa)\dn(\phi_0,\kappa)
\end{equation}
with \(\sn\) and \(\dn\) are also Jacobian elliptic functions.
It is simple to see from (\ref{rel}) that \(\kappa\) is the
integral of motion related to the Hamiltonian
\begin{equation}\label{scH}
\kappa=\sqrt{\lambda H/\omega^4},
\end{equation}
while the initial phase is
\begin{equation}\label{phi}
\phi_0=\cn^{-1}[\omega z/\sqrt{2H}].
\end{equation}
It also follows from (\ref{rel}) that for \(p_0>0\), the initial phase
changes in the range
3\(K(\kappa)<\phi_0<4K(\kappa)\) (or \(-K(\kappa)<\phi_0<0\)), where
\(K(\kappa)\) is the elliptic integral of the first kind.
The modulus is restricted by \(0\leq\kappa^2< 0.5\), and the relations
between the constants of integration and the initial conditions
are expressed by the single--valued functions.

Inserting (\ref{sol2}) in (\ref{S}), and using the integrals (312.02),
(312.04), (361.02) of the Ref. \cite{byrd} and the formulas for the elliptic
integral of the second kind \cite{mizrahi}
\[E(\phi)-E(\phi')=E(\phi-\phi')-\kappa^2\sn(\phi)\sn(\phi')\sn(\phi-\phi'),
\]
we obtain the following expression for the action \(S\)
\begin{eqnarray}\label{act1}
S= \frac{-2\omega^2\Omega}{3\lambda}
E(\Omega r)+\Omega^4r(1-\kappa^2)(2-3\kappa^2)/3\lambda
+\frac{2\omega^2\Omega\kappa^2}{3\lambda}\times \nonumber \\
\left\{\sn(\phi_0)\sn(\phi)\sn(\Omega r)
+\frac{\Omega^2}{\omega^2}[\sn(\phi_0)\cn(\phi_0)\dn(\phi_0)
-\sn(\phi)\cn(\phi)\dn(\phi)]\right\}
\end{eqnarray}
where
\[\phi=\Omega r+\phi_0,~~~\Omega=\frac{\omega}{\sqrt{1-2\kappa^2}}. \]
The following notations  \(E(x)\equiv E(x,\kappa) \)
and \(\sn(x)\equiv\sn(x,\kappa) \) (the same for \(\cn,\dn\)) are used.

\subsection{``Shelves'' in the small \(\kappa\) approximation}

{\em The small \(\kappa\) approximation.}
The expression for the action \(S\)
can be simplified. Since \(\kappa^2<0.5\), one can use the
small--\(\kappa\)--approximation for the elliptic integrals.
Using the following definition for the elliptic integral \cite{abram}
\[E(x,\kappa)\equiv E(x)=x-\kappa^2\int_0^x\sn(x')dx' \]
and the approximation
\(\sn(x)\approx \sin(x) \),
we obtain approximately that
\begin{equation}\label{AA1}
E(x)\approx x-x\kappa^2/2-\kappa^2\sin(2x)/4.
\end{equation}
Inserting(\ref{AA1}) in (\ref{act1}), and then combining  the first
two terms, we obtain, after doing small algebra, the following expression
for the action
\begin{equation}\label{act2}
S(\kappa)\approx\frac{\omega^4 r\kappa^4}{3\lambda}-
\frac{\omega^3\kappa^2}{2\lambda}\left[\sin(2\phi)-\sin(2\phi_0)\right],
\end{equation}
where the nonlinear frequency is now \(\Omega\approx \omega(1+\kappa^2)\).
It also follows in this approximation that the relation (\ref{rel})
between the initial momentum \(p_0\) and the modulus \(\kappa\)
is simplified
\begin{equation}\label{rel2}
p_0\approx g\kappa,
\end{equation}
where \(g=-\omega\sqrt{2/\lambda}\sin\phi_0 \), and
\(-\pi/2\leq\phi_0<0\).
The dependence of ray travel times on the initial momentum \(T(p_0)\) in
SVZ coincides up to some constant multiplier with the dependence of the
action
on the modulus \(S(\kappa)\) in (\ref{act2}).

{\em ``Shelves.''}
It follows that the action in the form (\ref{act2}) consists of two
terms. The first one is the dominant (proportional to \(\omega
r\gg 1\)) monotonic growing function in \(\kappa\). The second one is the
small but fast oscillating term with a large frequency (proportional to
\(\omega r\gg 1\)). Such a combination of these two terms ensures
the monotonic growth of the function in general, but at the same time
the extrema equation \(\partial S/\partial\kappa=0\) has solutions.
These solutions can be simply obtained,
{\em e.g.} for \(\phi_0=0\). The extremum points condition
gives, in the same limit \(\omega r\gg 1\), the following solutions for
\(\kappa\)
\begin{equation}\label{extrem}
2\phi=2\Omega(\kappa) r+2\phi_0=\pm\arccos(2/3)+2\pi m+O(1/\omega r)
\equiv\phi_m^{\pm},
\end{equation}
where \(m>\omega r/\pi\) are integer numbers and \(O(1/\omega r)\)
means neglected terms of the order of \(1/\omega r\ll 1\).

Therefore, there are regions between extrema points
\((\phi_m^{-},\phi_m^{+})\) with the same
number \(m\) where the derivatives are negative,
\(\partial S/\partial\kappa<0\). It follows that, in a range of
 \(\Delta\kappa =\Delta_{-}\approx\pi/8\omega r\kappa\), the action
\(S\) decreases by \(\Delta S_{-}\) (see Appendix A). These
regions alternate with regions of growth, where
\(\partial S/\partial\kappa>0\). Between extremum points
\((\phi_m^{+},\phi_{m+1}^{-})\) on the range of \(\Delta\kappa=
\Delta_{+}=3\Delta_{-}\) the action changes as
\begin{equation}\label{AApm}
\Delta S_{+}=9\Delta S_{-}
\end{equation}
Therefore the growth of the action is stronger (by 9 times) than the
decrease that leads to the step-like behavior of the action as a
function of \(\kappa\).
This step--like function (see Figs. 1 and 3) has horizontal parts
called ``shelves'' in \cite{SVZ02}.

An important feature of ``shelves'' is a large number of Fourier
components in the Fourier transformation of the oscillating term in
(\ref{act2}) (see Appendix B).
It is shown in Appendix C that the average number of ``harmonics''
of the Fourier transformation is
\begin{equation}\label{ApC}
\langle D_{s}\rangle\approx\omega r\gg 1
\end{equation}
One can see, in the insert of Fig. 3, a large number of the Fourier
amplitudes.

\section{Travel time distribution for chaotic rays}

In contrast to the regular dynamics, the arrival times of chaotic rays
are not uniquely defined  functions of the initial conditions, which is a
simple result of the energy \(H\) (\ref{Ham}) or the modulus \(\kappa\)
being no more the integrals of motion in the
chaotic case. This means that many initial conditions can contribute to
the same arrival time (as it is seen in Fig. 2). Wave dynamics leads
to the wave superposition with different probabilities for
different arrival times.
Obvious maxima of the travel times distribution  are seen in Fig. 2.
To explain this phenomenon, we will use the analytical solution for the
unperturbed ray dynamics, while chaotic dynamics is modeled by a
randomization of the initial phase \(\phi_0\) or by a variety of different
sources with random phases \(\phi_0\) uniformly distributed in the
interval \((-\pi/2,0)\).

{\em An integrable case.}
The probability of finding a particle with the range \(r\) and the depth
\(z\) is defined by a solution of the parabolic equation (\ref{Scheq})
with an amplitude (\ref{A}). Therefore these amplitudes define the
probability distribution for different \(S(\kappa)\) with the same fixed
\(r\) by (\ref{A})
\begin{equation}\label{A2}
|A(r,z)|^2\propto |{\prt z}/{\prt p_0}|^{-1}.
\end{equation}
Taking into account the solution \(z(r)\) in (\ref{sol2}) and the
relation \(p_0(\kappa)\) in (\ref{rel2}), we obtain in the
small--\(\kappa\)--approximation
\begin{eqnarray}\label{dzdp0}
&\prt z/\prt p_0=(\prt\kappa/\prt p_0)\cdot(\prt z/\prt\kappa)
+(\prt\phi_0/\prt p_0)\cdot(\prt z/\prt\phi_0)  \nonumber \\
&\approx [4\omega
r\kappa^2\cos\phi_0\sin\phi-2\cos(\phi+\phi_0)]/\sin(2\phi_0)
\end{eqnarray}
In the limit \(\omega r\gg 1\),
the main contribution to the derivative (\ref{dzdp0}) is due to the
linear term \(\phi\sim\omega r\). Therefore the evaluation of the
probability for the asymptotically large times is
\begin{equation}
\label{A3}
|A|^2\approx\frac{1}{2\omega r\kappa^2}
\left|\frac{\sin(\phi_0)}{\sin\phi}\right|.
\end{equation}
The maxima of this probability correspond to zeroes of the
denominator, which can be found from the following equation
\begin{equation}\label{kapn}
\phi(\kappa=\kappa_n)=\phi_0+\Omega(\kappa_n) r=\phi_0+\omega
r(1+\kappa_n^2)=
\pi n,~~~~~n=0, 1, 2,\dots
\end{equation}
For the fixed \(\omega r\) and \(\phi_0\), the solutions of (\ref{kapn})
\(\kappa=\kappa_n\) determine the actions \(S_n=S(\kappa_n)\) where the
maxima of probability take place for the integrable case.

{\em Ray chaos.}
For the chaotic case the energy \(H\) or the modulus \(\kappa\)
are no longer the integrals of motion. Therefore the rays with different
initial conditions \(\kappa,\phi_0\) can contribute to the same
arrival time \(S\) with different probabilities. In our phenomenological
approach, it is convenient, as was mentioned above, to model the
chaotic dynamics
by a variety of initial conditions with random phases \(\phi_0\) uniformly
distributed in the interval \((-\pi/2,0)\).
Therefore the averaged probability is a superposition of probabilities
with all initial phases. It reads
\begin{equation}\label{meanA}
\lgl |A|^2\rgl=\frac{2}{\pi}\int_{-\pi/2}^{0}|A|^2d\phi_0=
\frac{\pm 1}{\omega r\kappa^2}\left(\sin(\Omega r)/8-
(1/\pi)\cos(\Omega r)\ln\left[-\tan(\Omega r)\right]\right),
\end{equation}
where the signs \(\pm\) are due to the modulus function in (\ref{A3}) and
(\ref{meanA}), and \((+)\) sign stands for
\(-\pi/2<(\Omega r,~\mbox{mod}~2\pi) <0\), while
\((-)\) sign is taken for \(\pi/2<(\Omega r,~ \mbox{mod}~ 2\pi)<\pi\).
The maxima of the mean probability are
\begin{equation}\label{maxima}
\Omega(\kappa_n) r=\omega r(1+\kappa_n^2)=\pi n,
\end{equation}
that coincides with (\ref{kapn}) for \(\phi_0=0\). It follows from
(\ref{kapn}) and
(\ref{meanA}) that rays with different \(\kappa\) are clustered by the
index \(n\) numbering the maxima. For all values of \(\phi_0\)
one always finds a value of \(\kappa\) which corresponds to
the maxima conditions with the same \(n\). It also follows that
all other values of \(\kappa\) which do not correspond to the maxima
conditions ``carry'' the same index \(n\) if their action \(S\) is
close to the maximum value \(S(\kappa_n)\).
This phenomenon of the ray clustering can be a possible explanation of
the {\em ID number} for rays \cite{SVZ02}.

\section{Conclusion.}

It should be admitted that in the framework of this simple analysis
of the solution of the quartic oscillator, we are able to describe fairly
well the step--like behaviour of the arrival paths as a function of the initial
momentum. This step--like behaviour is known as ``shelves'' \cite{SVZ02}.

For the chaotic behaviour of rays, we constructed a phenomenological
model and presented only qualitative explanations of the nonuniform
distribution of the arrival paths as a function of the initial
momentum. The maxima of this distribution are explained in the framework
of the integrable model. Such a kind of consideration corresponds to a
so--called linear response approximation.

This work was supported by the U.S. Navy Grant N00014-97-1-0426.
We thank A. Pitt for her help in preparing  this paper.

\section*{Appendix A}
\def\theequation{A. \arabic{equation}}
\setcounter{equation}{0}
The extremum points condition
gives in the same limit \(\omega r\gg 1\) the following solutions for
\(\kappa\)
\begin{equation}\label{extremA}
2\phi=2\Omega(\kappa) r+2\phi_0=\pm\arccos(2/3)+2\pi m\equiv\phi_m^{\pm},
\end{equation}
where \(m>\omega r/\pi\) are integers numbers.
The phases \(\phi_m^{+} \) stand for the minima of S with
\begin{equation}\label{d2Sd2k1}
\partial^2S(\phi_m^{+}/\partial\kappa^2\equiv S_{m,+}^{\prime\prime}=
8\sqrt{5}\omega^5\kappa^4r^2 >0,
\end{equation}
while \(\phi_m^{-} \) define the maxima of the action,
\begin{equation}\label{d2Sd2k2}
\partial^2S(\phi_m^{-}/\partial\kappa^2\equiv S_{m,-}^{\prime\prime}=
-8\sqrt{5}\omega^5\kappa^4r^2.
\end{equation}
It is simple to see that the regions on \(\kappa\) between any adjoint
extrema are very small. Indeed, the width of the region where the action
decreases, \(\Delta_{-}\) is determined from (\ref{extrem})
\[
\phi_m^{+}-\phi_m^{-}=2\omega r[(\kappa+\Delta_{-})^2-\kappa^2]=\pi/2,
\]
where we took approximately that \(\arccos(2/3)\approx\pm\pi/4\).
From where we obtain that
\begin{equation}\label{Dm}
\Delta_{-}\approx\pi/8\omega r\kappa.
\end{equation}
Analogously, from \(\phi_{m+1}^{-}-\phi_m^{+}=3\pi/2\) we obtain
that the width of the region where \(S\) increases is
\begin{equation}\label{Dp}
\Delta_{+}=3\Delta_{-}.
\end{equation}
Since \(\Delta_{\pm}\ll 1\), we can define both a growth
\(\Delta S_{+}\) and a decrease \(\Delta S_{-}\) of the action in
corresponding
regions between adjoined extremal points in the linear approximation.
Expanding the first derivative \(\partial S/\partial\kappa\) near
every extremal point, we obtain for \(\Delta S_{-}\)
\[
\Delta S_{-}=\int_0^{\Delta_{-}/2}S_{m,-}^{\prime\prime}xdx
+\int_0^{\Delta_{-}/2}S_{m,+}^{\prime\prime}(-x)dx.
\]
Inserting (\ref{d2Sd2k1}) and (\ref{Dm})  in the integration, we obtain
that
\begin{equation}\label{DSmin}
\Delta S_{-}=-\pi^2\sqrt{5}\omega^3\kappa^2/16.
\end{equation}
Carrying out the same for \(\Delta S_{+}\) we obtain
\begin{equation}\label{DSmax}
\Delta S_{+}=9|\Delta S_{-}|.
\end{equation}

\section*{Appendix B}
\def\theequation{B. \arabic{equation}}
\setcounter{equation}{0}
Let us rewrite the oscillating term in the form
\begin{equation}\label{expan}
\sin(2\phi)=\sin(2\omega r+2\phi_0)\cos(\omega r\kappa^2)+\cos(2\omega
r+2\phi_0)\sin(\omega r\kappa^2).
\end{equation}
For simplicity we consider \(\kappa\in[0,1]\) by rescaling
\(2\kappa^2\rightarrow\kappa^2\) that does not lead to any
errors in the analysis.
Since the region of definition of \(\sin( 2\phi)\) is restricted by this
segment, it is not difficult to show that the coefficients of the Fourier
transformation \(f^C(s),f^S(s)\) are determined by
the Fresnel integrals \(C(s),S(s)\) \cite{gradshtein,abram}:
\begin{equation}\label{FS}
f^C(s)=\int_0^1d\kappa\sin(2\phi)\cos(2\pi s\kappa),~~~
f^S(s)=\int_0^1d\kappa\sin(2\phi)\sin(2\pi s\kappa).
\end{equation}
Carrying out the variable change \(x=\omega r\kappa\) and considering
that \(\omega r\gg 1\) we take the upper limit
to \(\infty\). Then we have for (\ref{FS}) the following four integrals
which determine the coefficients \(f^C(s),f^S(s)\)
\[\int_0^1d\kappa\sin(\omega r\kappa^2)\sin(s\kappa)\rightarrow
\frac{1}{\sqrt{\omega r}}
\int_0^{\infty}dx\sin(x^2/\omega r)\sin(\frac{s}{\omega r}x)
\]
and it gives
\[
\sqrt{\frac{\pi}{2\omega r}}\left\{\cos\frac{s^2}{4\omega r}
C\left(\frac{s}{2\sqrt{\omega r}}\right)+
\sin\frac{s^2}{4\omega r}S\left(\frac{s}{2\sqrt{\omega r}}\right)\right\},
\]
Analogously we  obtain for the rest of integrals
\[
\int_0^1d\kappa\sin(\omega r\kappa^2)\cos(s\kappa)\approx
\sqrt{\frac{\pi}{8\omega r}}\left\{
\cos\frac{s^2}{4\omega r}-\sin\frac{s^2}{4\omega r}\right\},
\]
\[
\int_0^1d\kappa\cos(\omega r\kappa^2)\cos(s\kappa)\approx
\sqrt{\frac{\pi}{8\omega r}}\left\{
\cos\frac{s^2}{4\omega r} +\sin\frac{s^2}{4\omega r}\right\},
\]
\[
\int_0^1d\kappa\cos(\omega r\kappa^2)\sin(s\kappa)
\]\[
\approx
\sqrt{\frac{\pi}{2\omega r}}\left\{
\sin\frac{s^2}{4\omega r}C\left(\frac{s}{2\sqrt{\omega r}}\right)
-\cos\frac{s^2}{4\omega r}S\left(\frac{s}{2\sqrt{\omega
r}}\right)\right\}.
\]
Keeping \(\Delta s\Delta x>const\), we obtain that there are of the order
of \(\omega r\gg 1\) components with the amplitudes \(\sim 1/\sqrt{\omega r}\)
contributed to the Fourier transformation.

\section*{Appendix C}
\def\theequation{C. \arabic{equation}}
\setcounter{equation}{0}
The oscillating part of the action \(S\) has a complete oscillation
between points \((\phi_m^{\pm},\phi_{m+1}^{\pm})\) that corresponds to
the range on \(\kappa\) or a quasi--period equaled to
\begin{equation}\label{CC1}
D_{\kappa}=\Delta_{+}+\Delta_{-}=4\Delta_{-}.
\end{equation}
Hence, taking into account (\ref{Dm}), we obtain that the number of
harmonics in the Fourier transformation is
\begin{equation}\label{CC2}
D_s=2\pi/D_{\kappa}=4\omega r\kappa.
\end{equation}
Since \(0<\kappa^2<0.5\), the averaging  of (\ref{CC2}) gives
\begin{equation}\label{CC3}
\langle D_s\rangle=\omega r.
\end{equation}
It should be stressed that this estimate is approximate and gives only the
order of \(D_s\). The exact theorem on the uncertainty conditions
(see for example \cite{papoulis}) ensures only that
\(\langle D_s\rangle>\omega r/\sqrt{8\pi}\).

\newpage

\section*{Figure captions}

\noindent Fig. 1.
The ray travel time as a function of starting momentum for two waveguides
with the sound speed profiles \( c_{1}(z)=c_{01}+az^{2}+bz^{4} \)
(curves \( a \), \( b \), and \( c \)), and \( c_{2}(z)=c_{02}-\gamma z \)
(curves \( d \), \( e \), and \( f \)). Parameters: \( c_{01}=1.49 \)
km~s\( ^{-1} \), \( a=1. \) km\( ^{-1} \)s\( ^{-1} \), \( b=1. \)
km\( ^{-3} \) s\( ^{-1} \), \( c_{02}=1.4465 \)
km~s\( ^{-1} \), \( \gamma =0.0435 \) s\( ^{-1} \). It has been assumed
that  the waveguide with \( c_{1}(z) \) has no boundaries, while \( c_{2}(z) \)
has a reflecting surface at \( z=0 \). The travel time at each curve is
estimated from the arrival of the ray with \( p_{0}=0. \) Different curves
present rays escaping point sources located at depths: \( 0 \) km (\( a \)),
\( 0.5 \)  km (\( b \)), \( 1 \) km (\( c \)), \( 0 \) km (\( d \)), \( 1 \) km
(\( e \)), and \( 2 \) km (\( f \)).
[from Ref. \cite{SVZ02}]

\noindent Fig. 2.
The ray travel time versus starting momentum in the unperturbed
(thick solid lines) and perturbed (points) waveguides at the range of 4500
km and for the point source set at a depth of 2.5 km.
[from Ref. \cite{SVZ02}]

\noindent Fig. 3.
The ray travel time (action \(S\) versus the modulus \(\kappa\) for
Eq. (\ref{act2}), where \(\phi_0=-\pi/4,~\omega=1,~r=355.4,~\lambda=1.2\).
The insert is  the amplitudes \(f(s)\) vs \(s\) of the discrete Fourier
transformation for the oscillating part of the action \(S\) (\ref{FS}).

\begin{figure}[htbp]
\epsfxsize=0.7\textwidth
\centerline{%
\epsffile{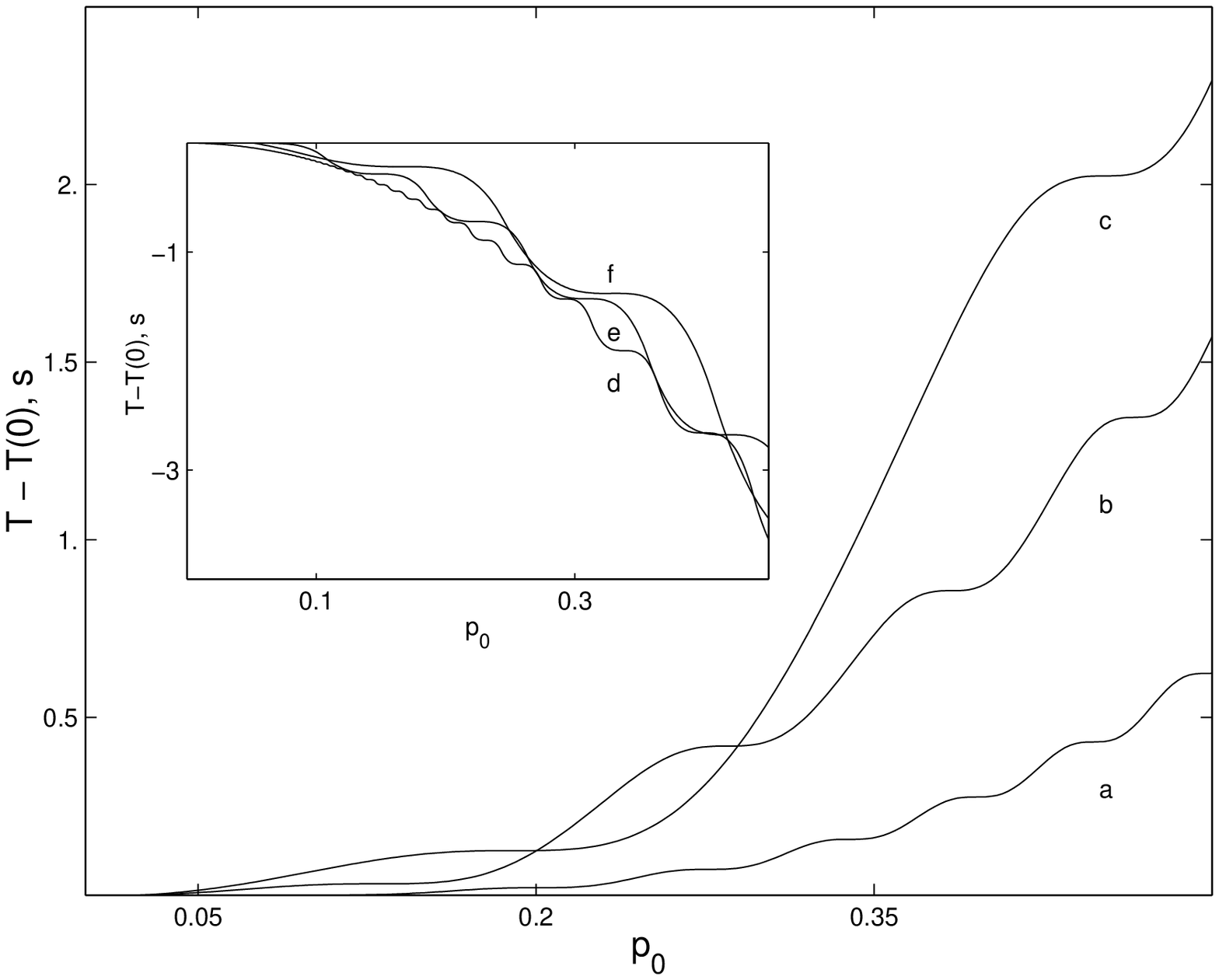}%
}
\caption{}
\end{figure}

\begin{figure}[htbp]
\epsfxsize=0.7\textwidth
\centerline{%
\epsffile{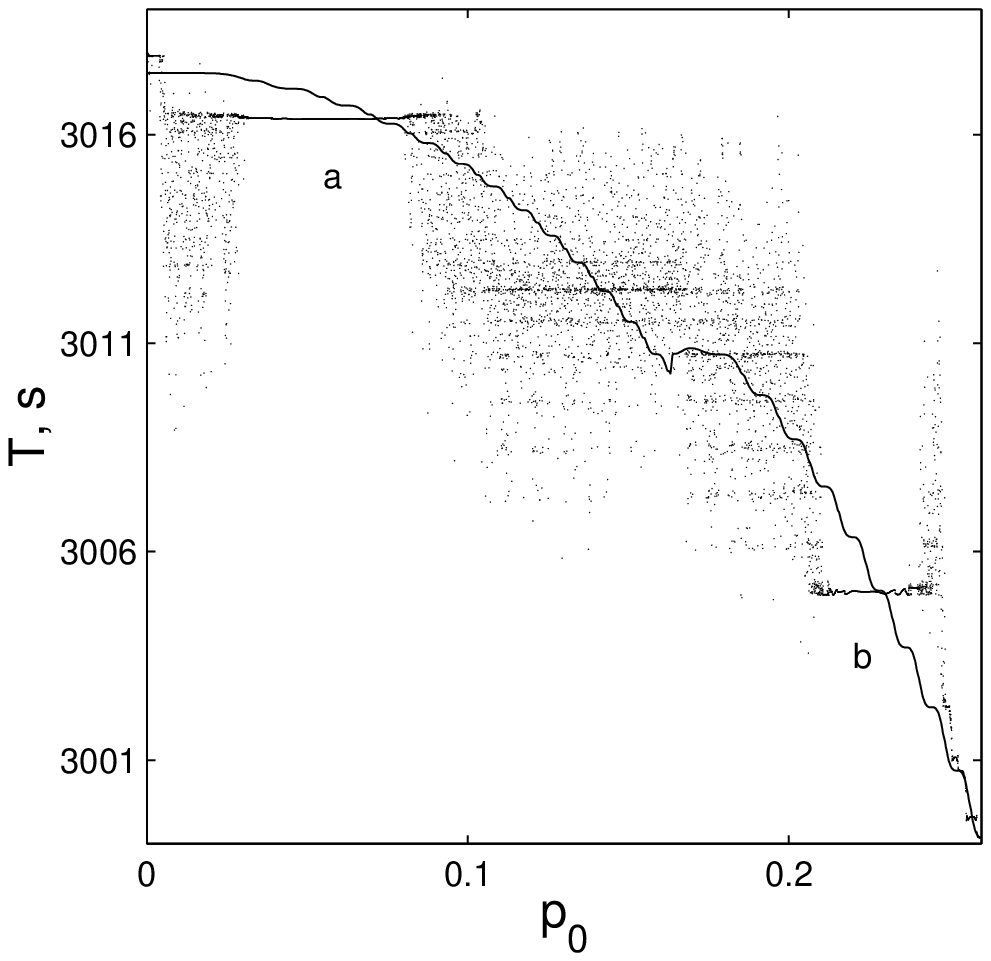}%
}
\caption{}
\end{figure}

\begin{figure}[htbp]
\epsfxsize=0.7\textwidth
\centerline{%
\epsffile{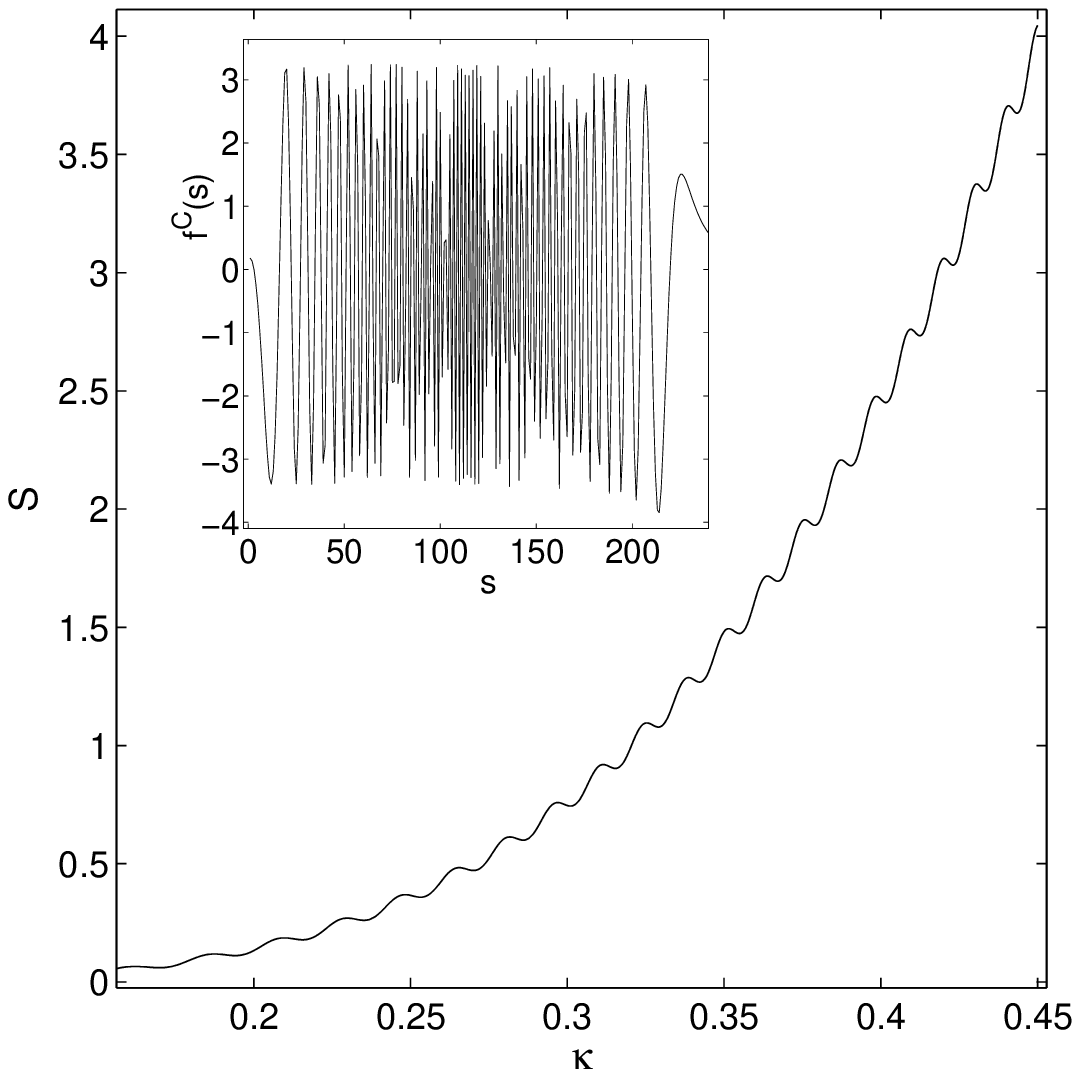}
}
\caption{}
\end{figure}

\end{document}